\documentclass[twocolumn,prl,showpacs,amsmath,amssymb]{revtex4}

\usepackage{graphicx}
\usepackage{dcolumn}
\usepackage{bm,color}

\begin{document}

\preprint{APS/}

\title{Suppression of Charge Equilibration leading to the Synthesis of Exotic Nuclei}

\author{Yoritaka Iwata$^{1}$}
\author{Takaharu Otsuka$^{2,3}$}%
 \author{Joachim A. Maruhn$^{4}$}
\author{Naoyuki Itagaki$^{2}$}
 \affiliation{$^{1}$GSI Helmholtzzentrum f\"ur Schwerionenforschung, D-64291 Darmstadt, Germany}
 \affiliation{$^{2}$Department of Physics, University of Tokyo, Hongo, Tokyo 113-0033, Japan}
 \affiliation{$^{3}$Center for Nuclear Study, University of Tokyo, Hongo, Tokyo 113-0033, Japan}
 \affiliation{$^{4}$Institut f\"ur Theoretische Physik, Universit\"at Frankfurt, D-60325 Frankfurt, Germany}

\date{\today}

\begin{abstract}
Charge equilibration between two colliding nuclei can take place in the early stage of heavy-ion collisions.  
A basic mechanism of charge equilibration is presented in terms of the
 extension of single-particle motion from one nucleus to
 the other, from which the upper
 energy-limit of the bombarding energy is introduced for
 significant charge equilibration at the early stage of the collision.
 The formula for this limit is presented, and is compared to
 various experimental data.  It is examined also by comparison to three-dimensional time-dependent density functional calculations.
The suppression of charge equilibration, which appears in collisions at the energies beyond the upper energy-limit, gives rise to remarkable effects on the synthesis of exotic nuclei with extreme proton-neutron asymmetry.
\end{abstract}

\pacs{21.10.Ft, 25.70.-z, 25.70.Hi}

\maketitle

Charge equilibration is a rapid process during the early stage of heavy-ion collisions with a time scale of $10^{-22}$~sec  
(for a review, see \cite{84Freiesleben}).
Despite many theoretical attempts, its mechanism has remained an open problem, where only the relation with the isovector giant dipole resonance (iv-GDR) was discussed in certain cases (for example, 
see \cite{Berlanger79, Hernandez81, BonN81, SurSch89, 01Baran, Sim01, Sim07}). 
The charge equilibration is quite important, because it naturally prevents the synthesis of exotic nuclei with extreme proton-neutron asymmetry. 
As there are and will be the 3rd generation RI-beam facilities, it is an urgent question whether such synthesis can be enhanced with higher beam energies or not.

In this Letter, we first point out that there is an upper limit of the bombarding energy for the fast and significant charge equilibration in the initial stage of the collision.
Microscopic three-dimensional (time-dependent) density functional calculations are systematically performed for the purpose of examination.
We actually employ Skyrme time-dependent Hartree-Fock (TDHF) theory, which is a rather unique presently feasible method for the treatment of non-perturbative processes such as multi-nucleon transfers in a realistic framework.

\begin{figure*} 
\includegraphics[width=165mm]{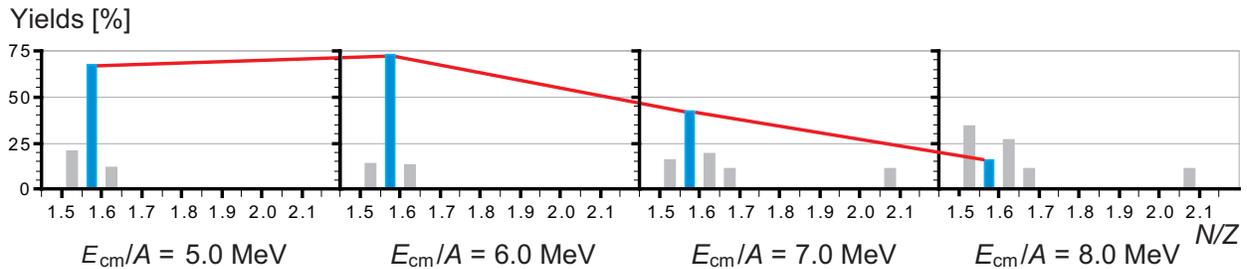}
\caption{\label{fig1} Yield distribution of final fragments as a
 function of their $N/Z$ ratios for the collisions of $^{208}$Pb + $^{132}$Sn ($N/Z$ ratio is discretized by 0.05) based on TDHF calculation (SLy4d).
Four cases with different $E_{cm}/A$ values are presented.
Columns corresponding to the equilibrium value of $N/Z$=1.58 are colored in blue, and connected by red lines.
For reference, the $N/Z$ ratios for $^{208}$Pb and $^{132}$Sn are 1.54 and 1.64, respectively.  }
\end{figure*}

The suppression of charge equilibration brings about a favorable
situation for the synthesis of exotic nuclei.  
The evaluation of the upper limit thus can have crucial significance for experiments on nuclei with extreme proton-neutron asymmetry.
We shall consider the collision of a target nucleus with mass number
$A_1$, neutron (proton) number $N_1$ ($Z_1$), with a projectile 
nucleus with $A_2$, $N_2$, and $Z_2$.
The total mass, neutron, and proton numbers are denoted by $A$, $N$, $Z$, respectively.
We look at this problem from an intuitive and basic viewpoint.  
We begin with a picture that the charge equilibration takes place as wave functions of nucleons propagate from their original nucleus to the other nucleus in the initial stage of the collision.
Namely, the regime of individual single-particle motion spreads out following the lowering of potential barrier between the two nuclei after the touching.  
Because this spreading occurs as a consequence of unblocked 
single-particle motion, the process can be very fast;  
a particle travels into the other side within $\sim 10^{-22}$ sec
at a quarter of light velocity, which roughly corresponds to the Fermi energy of normal nuclear matter. 
On the other hand, this propagation can be prohibited by the Pauli
principle, while the neutron (or proton) excess weakens this effect.
Another important factor is the difference of velocities of the two nuclei.  
Because this spreading needs a certain time, it does not lead to charge
equilibration in the initial stage if the relative velocity of the
colliding nuclei is too large.  
In the charge equilibration, protons and neutrons from both nuclei, 
particularly those near the Fermi levels, are mixed within a certain time after the touching.  

We can introduce an ansatz that, in order for the fast charge equilibration to occur, the relative velocity $v_{r}$ of the two nuclei at the collision must be below the velocities corresponding to the proton or neutron Fermi momenta of both nuclei.  
By denoting the minimum of these four velocities as $v_{F}^{\rm~min}$, the present effect can be summarized by the statement that the upper limit of the bombarding energy for charge equilibration is determined by $v_{r}=v_{F}^{\rm~min}$.
The upper limit of the energy in the laboratory frame for charge equilibration is expressed as the sum of the kinetic energy for velocity $v_{F}^{\rm~min}$ and the Coulomb energy at touching:
\begin{eqnarray} 
\label{eq1}
&& E_{{\rm CE}, lab}/A {\rm [MeV]}  \vspace{2.5mm} \nonumber \\
&& = \frac{\hbar^2 (3 \pi^2 \rho_{\rm min})^{2/3}}{2m} 
+ \frac{e^2 Z_1 Z_2}{4 \pi \epsilon_0 r_0} \frac{A_1 + A_2} {A_1 A_2 (A_1^{1/3}+A_2^{1/3})}, 
\end{eqnarray}
\begin{eqnarray} 
\label{eq1d}
&\rho_{\rm min} = 
{\displaystyle \min_{i}} \big(\rho_{ni},\rho_{zi} \big) 
\hspace{40mm} \vspace{2.5mm} \nonumber \\
& = {\displaystyle \min_{i}} \left( \frac{N_{i}  \left( \frac{4 \pi r_0}{3}  A_i^{1/3} \right)^{-1} }{(1- 3{\bar \epsilon})(1+{\bar \delta})}, \frac{Z_{i}  \left( \frac{4 \pi r_0 }{3} A_i^{1/3}\right)^{-1} }{(1- 3{\bar \epsilon})(1-{\bar \delta})}\right), 
\end{eqnarray}
where $m$, $e$, $\epsilon_0$, and $r_0$ are the nucleon mass, 
the charge unit, the vacuum permittivity, 
and the usual nuclear radius parameter (1.2~fm), respectively. 
Here we express the minimum velocity by the corresponding minimum
of the proton or neutron density of the two nuclei using the formula
$ v_{F}^{\rm~min} = \hbar(3\pi^2 \rho_{\rm min})^{1/3} /m$.
In Eq.~(\ref{eq1d}), $i=1,2$ distinguishes the two initial nuclei, and 
${\bar \epsilon}$ and ${\bar \delta}$ (functions of $A_i$) are introduced based on the droplet model \cite{MyerSwia69,Myer69} to take into account the effect of neutron and proton skins.

\begin{table}  
\begin{center}
\caption{$E_{cm}/A$ values [MeV] of
 the upper limit of charge equilibration obtained by TDHF 
 calculations with SLy4d and SkM* parameters, compared to those 
 by the proposed formula, eq.~(\ref{eq1}). 
For reference, the values obtained by the Fermi
 gas model with standard parameters are shown.} 
\vspace{2.5mm}
\protect
 \label{table1}
\begin{tabular}{|c|l|r|r|r|r|} \hline & Collision & \quad TDHF   &  \quad  TDHF   & Eq. (\ref{eq1}) & Fermi \\ 
& & \quad (SLy4d)   &  \quad  (SkM*)   &  & gas \\ \hline (i) &
$^{208}$Pb + $^{238}$U   & 6.5$\pm$0.5 &  6.5$\pm$0.5 &  6.91 &  9.46  \\
 (ii) & $^{208}$Pb + $^{132}$Xe  & 6.5$\pm$0.5 & 6.5$\pm$0.5 &  6.50 &  9.03 \\
 (iii) & $^{208}$Pb + $^{132}$Sn  & 6.5$\pm$0.5 &  6.5$\pm$0.5 &  6.36 & 9.03  \\
 (iv) & $^{208}$Pb + $^{\; \> 40}$Ca  &  3.5$\pm$0.5 &   3.5$\pm$0.5 & 3.66 &  5.14 \\
 (v) & $^{208}$Pb + $^{\; \> 24}$Mg  &  2.5$\pm$0.5 & 2.5$\pm$0.5  & 2.36 &  3.52  \\  
 (vi) & $^{208}$Pb + $^{\; \> 24}$O   &  2.5$\pm$0.5 &  2.5$\pm$0.5 & 2.18 &  3.52 \\    
 (vii) & $^{208}$Pb + $^{\; \> 16}$O   & 1.5$\pm$0.5 &  1.5$\pm$0.5 & 1.75 &  2.50 \\  
 (viii) & $^{208}$Pb + $^{\; \> \; \> 4}$He & $<$ 1.0 ~ &    $<$ 1.0 ~ &  0.48  &  0.70 \\
 (ix) & $^{ \> 24}$Mg  + $^{\; \> 24}$O  &  5.5$\pm$1.0  & 5.5$\pm$1.0 & 5.99 &  9.5 \\ 
\hline
\end{tabular}
\end{center}
\end{table} 

In order to confirm the validity of this picture, we perform three-dimensional TDHF calculations systematically.
Many TDHF events were obtained with different values of impact parameter being incremented by 0.25 fm.
These events are summed over impact parameters with weights of geometric cross section, and elastic cases are discarded.
By collecting TDHF events, we sort them in terms of $N/Z$ ratio with $N/Z$ being discretized into bins of width 0.05.
Figure \ref{fig1} shows the yield distribution of final fragments for $^{208}$Pb + $^{132}$Sn reaction.
Going from low to high $E_{cm}$ (total kinetic energy in the
center-of-mass frame) of the collision, the peak energy of the yield distribution as a function of $N/Z$ is almost constant at the beginning, and is shifted later with the lowering of peak height.
A clear decrease of the yield of charge-equilibrated fragments for 
$E_{cm}/A$ $\geq$ 7.0 MeV is noticed.
On the other hand, very neutron-rich nuclei with $N/Z \sim 2.0$ simultaneously start to be produced.
By taking $E_1$ as the energy at which the peak height starts to be
lowered (6.0 MeV in this case), and $E_2$ as the energy at which the
yield of the equilibrated $N/Z$ becomes about 50~$\%$ (7.0 MeV in this
case), the upper energy-limit is defined in TDHF as 
$E_{\rm CE} = (E_1 + E_2)/2$ (6.5 MeV in this case). 
The uncertainty from the energy bin value is $(E_2 - E_1)/2$ (0.5 MeV in this case).
Such TDHF results are summarized in the third and forth columns of Table \ref{table1}.
TDHF calculations with two different parameter-sets result in the same upper energy-limit. 
Note that a large value is obtained for $^{24}$Mg 
+ $^{24}$O, implying that a simple mass dependence (small values for reactions between light nuclei, and vice versa) cannot 
explain this, while the mass asymmetry plays a certain role.

\begin{figure} 
\includegraphics[width=75mm]{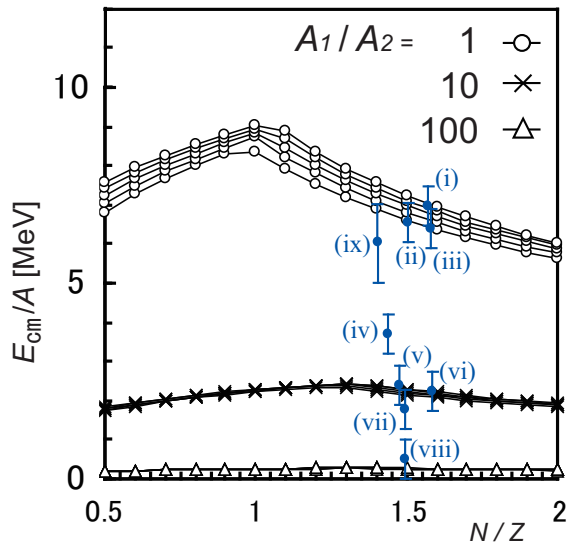}
\caption{\label{fig2} $N/Z$ dependence of the upper energy-limit of
 charge equilibration based on Eq. (\ref{eq1}) in the center-of-mass
 frame, where $A_1> A_2$ is assumed without loss of generality. 
Values with different total masses $A =$ 100, 200, 300, 400, and 500 are plotted for each $A_1/A_2$, which correspond to black lines from bottom to top in each group (values are too crowded to distinguish the total mass difference, for the cases of $A_1/ A_2$ = 10 and 100), and lines are drawn to guide eyes. 
Each TDHF calculation is shown as a blue bar, where the central points correspond to the value obtained by Eq. (\ref{eq1}) for each reaction, and Roman numbers distinguish reactions shown in Table \ref{table1}.}
\end{figure}

Let us have a comparison between the present upper-limit and the TDHF results.
Figure \ref{fig2} depicts how the upper limit given by Eq.~(\ref{eq1}) changes as functions of $N/Z$ and $A_1/A_2$ (or $A_2/A_1$).  
The upper energy-limit comes down significantly low for higher mass asymmetry, while it depends only weakly on the total mass.
Although charge equilibration can compete with Coulomb excitation
particularly in collisions involving nuclei with large $Z$, no evidence
of a major change due to large $Z$ is found in the results of
Eq. (\ref{eq1}) or those of TDHF calculations.
However, non-negligible decrease of the upper energy-limit due to the proton-neutron asymmetry is noticed.
The corresponding values obtained by the formula in Eq. (\ref{eq1}) is shown in the fifth column of Table \ref{table1}. 
For reference, the values obtained by the simple Fermi gas model ($k_F =
1.36$ fm$^{-1}$ for both protons and neutrons) are shown in its sixth column. 
By comparing the results of Eq. (\ref{eq1}) with the TDHF results shown in the third and forth columns, the agreement is remarkable, including a high value in the last row.
Comparison between the results of Eq.(\ref{eq1}) (blue circles) and the TDHF results (blue bars) is also made in Fig. \ref{fig2}. 
Consequently, the upper-limit of charge equilibration depends largely on the Fermi energy, and the proton-neutron asymmetry can contribute to the shift of the upper-limit, where the dependence of the upper-limit on the mass-asymmetry is remarkable.

Let us move on to the comparison to the experiments.
It can be seen that existing experimental data agrees with the present upper-limit formula.
For instance, the following experiments show charge equilibration:
$^{40}$Ar + $^{58}$Ni at $E_{lab}/A$=7.0 MeV \cite{75gatty}, and
$^{56}$Fe + $^{165}$Ho ($^{209}$Bi) at $E_{lab}/A$= 8.3 MeV \cite{79breuer}.
The following experiment shows the disappearance of charge equilibration: $^{112}$Sn + $^{124}$Sn at $E_{lab}/A$=50 MeV \cite{02Milazzo, 01Tan}. 
Recently, experiments: $^{124, 112}$Sn + $^{124, 112}$Sn at $E_{lab}/A$= 35 and 50 MeV respectively was performed at Michigan State University \cite{Tsang-priv}.
One remarkable result here is that the final fragments are not so close to charge equilibrium even in lower energy collisions with $E_{lab}/A$=35 MeV.
The disappearance of charge equilibration for this
experiment has not been explained. 
Because the upper limit of charge equilibration is calculated to be
$E_{lab}/A$=27.6 MeV from Eq. (\ref{eq1}), this
question can be understood now.

\begin{figure} 
\includegraphics[width=80mm]{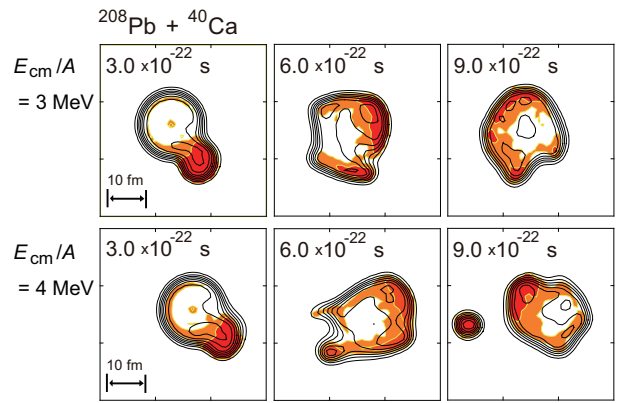}
\caption{\label{fig3} Real-time dynamics of charge distribution for $^{208}$Pb + $^{40}$Ca (SLy4d) is shown for a fixed impact parameter 7.5 fm.
The upper and the lower panels show cases with $E_{cm}/A$ = 3.0 MeV and 4.0 MeV, respectively.
For both cases, $^{208}$Pb is coming from the left, and $^{40}$Ca from the right. 
The colored regions show the distribution of charge (proton-rich part), and the contours being incremented by 0.02 fm$^{-3}$ show the density.
Domains with the proton-to-neutron density ratio greater than 0.78 and 0.72 are colored in red and yellow, respectively.}
\end{figure}

The suppression of charge equilibration contribute 
naturally to the production of exotic fragment; more
exotic nuclei far from the equilibrated $N/Z$ ratio are 
to be synthesized above the present upper-limit.  
Because the bombarding energies of the 3rd generation RI-beam facilities are sufficiently high to exceed the upper-limit, the present upper-limit ensures more production of further exotic isotopes by the latest and the future RI-beam facilities. 
In fact, in the experiment of Ref.~\cite{07Mocko}, the yield of exotic
fragments was increased simply by putting the beam energy higher than
the present upper-limit. 
It is also of 
much interest to explore the novel possibility of the synthesis of exotic nuclei by collisions including only $\beta$-stable nuclei.
Such possibilities have not attracted much attention, but
may emerge with various feasibilities of the production
of exotic spieces, if the experiment is set for energies beyond the
present upper-limit.  As an example, 
Fig.~\ref{fig3} shows the real-time dynamics of $^{208}$Pb + $^{40}$Ca
around the energy where the production of exotic nuclei
starts; the upper energy limit is 3.66 MeV (See
Table~\ref{table1}), fusion appears at $E_{cm}/A$ =
3.0 MeV (upper panels), and break-up is seen at 4.0 MeV (lower panels).
$\beta$-unstable fragments are emitted only for the bombarding energy
higher than the upper-limit; for instance, a small fragment in the lower-right panel 
of Fig.~\ref{fig3} is $^{62}$Mn (numbers of nucleon are rounded to be
integer), for which the stable isotope is $^{55}$Mn.  

\begin{figure} 
\includegraphics[width=70mm]{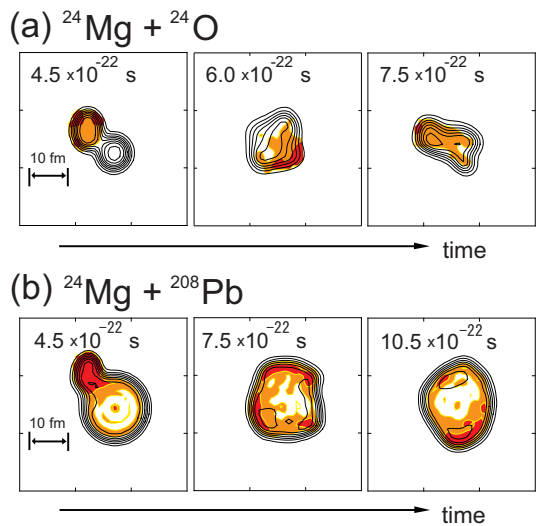}
\caption{\label{fig4} Real-time dynamics of charge distribution for $^{24}$Mg + $^{24}$O and $^{24}$Mg + $^{208}$Pb (SLy4d) are shown for $E_{cm}/A$ = 4.5 MeV and 2.0 MeV with the impact parameters 5.0 fm and 7.5 fm, respectively.
For both cases, $^{24}$Mg is incoming from the left, and the states are evolving into fusion.
The description manner is the same as Fig. \ref{fig3}, where parts with the proton to neutron density ratio larger than 1.00 (0.75) and 0.80 (0.70) are colored in red and yellow in case of $^{24}$Mg + $^{24}$O ($^{24}$Mg + $^{208}$Pb), respectively.}
\end{figure}

We now point out some interesting details of charge-equilibration
dynamics
for collisions below the present upper energy limit.
We shall first investigate it with a focus on the iv-GDR in a collision between light nuclei.
Figure \ref{fig4} (a) shows the time-evolution of charge distribution for $^{24}$O + $^{24}$Mg.
We see the appearance of the iv-GDR and the oscillation of well-localized charge (4.5, 6.0 $\times 10^{-22}$ sec).
The charge equilibration is synchronized with the iv-GDR.
This is, however, seen only for collisions between light nuclei.
Similar results are obtained in \cite{Berlanger79, Hernandez81, BonN81, SurSch89, 01Baran, Sim01, Sim07}. 
Next, we move on to collisions involving a heavier nucleus; e.g.  $^{24}$Mg + $^{208}$Pb, as shown in Fig. \ref{fig4} (b).
Charge-equilibration dynamics is not similar to the motion of the iv-GDR in this case, because no significant oscillations of the localized charge appear. 
Instead a radially layered structure of the composite nucleus is formed (7.5 $\times 10^{-22}$ sec), in which a relatively neutron-rich core appears in the center, a proton-rich layer surrounds it, and a neutron-rich skin is in the surface.
This seems to be due to the Coulomb repulsion, hence the radial distribution of charge is formed.
Here one can find analogous situation of isovector monopole excitation.
It means that the radial flow plays a prominent role in such low energy collisions.   
Similar layered structures are obtained in the TDHF calculations (with both Skyrme parameter-sets) listed in Table~\ref{table1} except for $^{24}$Mg + $^{24}$O.  
Consequently, charge equilibration, which is dependent on the mass, has been clarified to be related not only with the isovector giant dipole excitations, but also with the isovector giant monopole excitations.

Finally, let us comment on the diffusion-type mechanism towards charge equilibrium.
The mechanism of charge equilibration discussed in this Letter is valid at lower energies (lower than the upper energy-limit).
At much higher energies, collisions between nucleons and the diffusion contribute mostly to charge equilibration (for example, see \cite{03Shi,09Tsang}).
As the diffusion is a slow process, it is unlikely to attain the charge equilibrium within the initial stage.
Energy-dependent equilibration mechanisms of a similar kind have been reported in condensed matter physics (see \cite{57Landau,66Abel,75legget}), where the equilibration is fast for lower energies (temperatures), and quite slow for higher energies. 
Such a similarity between completely different physical systems is interesting.

In this Letter the mechanism of charge equilibration, in which
nucleons with the Fermi velocity play a primary role, has been
presented, and its validity is examined by comparison 
to virtually all existing relevant experimental data.
This concept has been further analyzed in terms of a three-dimensional
time-dependent density function formalism.  
The upper energy-limit of charge equilibration has a crucial impact 
on the nuclear synthesis of exotic nuclei.
Properties presented in this Letter will give a sound motivation for the production of further exotic isotopes by the latest and the future RI-beam facilities. 

This work was supported in part by EMMI project of the Helmholtz
Alliance, by JSPS Core-to-Core program EFES and by grant-in-aid for
Scientific Research (A) 20244022.
Y.I. thanks for the JSPS fellowship, and express his gratitude to
Dr. C. Simenel.  T.O. thanks Prof. P. Bonche for long-term valuable 
collaboration.


\begin{thebibliography}{1}
  \bibitem{84Freiesleben}
H.~Freiesleben and J.V.~Kratz, Phys. Rep. {\bf 106} (1984) 1. 

  \bibitem{Berlanger79}
M. Berlanger, A. Gobbi, F. Haneppe, U Lynen, C. Ng${\hat {\rm o}}$ and A. Olmi, Z. Phys. {\bf A 291} (1979) 133.

  \bibitem{Hernandez81}
E.S.~Hernandez, W.D.~Myers, J.~Randrup and B.~Remaud, Nucl. Phys. {\bf A 361} (1981) 483.

  \bibitem{BonN81}
P. Bonche and N. Ng${\hat {\rm o}}$, Phys. Lett. {\bf B 105} (1981) 17.

  \bibitem{SurSch89}
E. Suraud, M. Pi, and P. Schuck, Nucl. Phys. {\bf A 492} (1989) 294.

  \bibitem{01Baran}
V.~Baran, D.M.~Brink, M.~Colonna, and M. Di Toro, Phys. Rev. Lett. {\bf 87} (2007) 182501.

  \bibitem{Sim01}
C. Simenel, Ph. Chomaz, and G. de France, Phys. Rev. Lett. (2001) 2971.

  \bibitem{Sim07}
C. Simenel, Ph. Chomaz, and G. de France, Phys. Rev. {\bf C 76} (2007) 024609.

  \bibitem{MyerSwia69}
W.D.~Myers and W.J.~Swiatecki, Ann. Phys. {\bf 55} (1969) 395.

  \bibitem{Myer69}
W.D. Myers, Phys. Lett. {\bf 30B} (1969) 451. 

  \bibitem{75gatty}
B. Gatty, D.~Guerreau, M. Lefort, X. Tarrago and J. Galin, Nucl. Phys. {\bf A 253} (1975) 511.

 \bibitem{79breuer}
H. Breuer {\it et. al.}, Phys. Rev. Lett. {\bf 43} (1979) 191.

  \bibitem{02Milazzo}
P. M. Milazzo {\it et. al.}, Phys. Rev. {\bf C66} (2002) 021601.

  \bibitem{01Tan}
W.P.~Tan {\it et. al.}, Phys. Rev. {\bf C64} (2001) 051901.

  \bibitem{07Mocko}
M. Mocko {\it et. al.}, Phys. Rev. {\bf C76} (2007) 014609.

  \bibitem{Tsang-priv}
M.B.~Tsang {\it et. al.}, Private communication. 

  \bibitem{03Shi}
L. Shi and P. Danielewicz, Phys. Rev. {\bf C 68} (2003) 064604.

  \bibitem{09Tsang}
M.B.~Tsang {\it et. al.}, Phys. Rev. Lett. {\bf 102} (2009) 122701.

  \bibitem{57Landau}
L.D.~Landau and Zh. Eksperim, Soviet Phys. - JETP {\bf 5} (1957) 101.

  \bibitem{66Abel}
W.R.~Abel, A.C.~Anderson, and J.C.~Wheatley, Phys. Rev. Lett. {\bf 17} 2 (1966) 74.

 \bibitem{75legget}
A.J.~Leggett, Rev. Mod. Phys. {\bf 47} No.2 (1975) 331.

\end{thebibliography}
\end{document}